\documentclass[prx,twocolumn,superscriptaddress]{revtex4-2}
\usepackage{amsmath,amssymb,bm,mathrsfs,graphicx, braket, times,color}
\usepackage[colorlinks=true,citecolor=green!70!black]{hyperref}
\usepackage{soul}
\usepackage{xcolor}
\usepackage{colortbl}
\usepackage[makeroom]{cancel}
\usepackage{array,multirow,graphicx}
\usepackage{tikz}
\usetikzlibrary{shadings}
\usepackage{txfonts}  
\usepackage{txfontsb} 

\makeatletter 
\renewcommand{\section}{\@startsection{section}{1}{0mm}
  {-\baselineskip}{0.5\baselineskip}{\bf\leftline}}
\makeatother

\begin{document}

\title{Light-induced in-plane Rotation of Microobjects on Microfibers}

\author{Wei Lv}
\affiliation{Zhejiang University, Hangzhou 310027, Zhejiang Pronvince, China}
\affiliation{Key Laboratory of 3D Micro/Nano Fabrication and Characterization of Zhejiang Province, School of Engineering, Westlake University, 18 Shilongshan Road, Hangzhou 310024, Zhejiang Province, China}
\affiliation{Institute of Advanced Technology, Westlake Institute for Advanced Study, 18 Shilongshan Road, Hangzhou 310024, Zhejiang Province, China}

\author{Weiwei Tang}
\thanks{tangweiwei@westlake.edu.cn}
\affiliation{Key Laboratory of 3D Micro/Nano Fabrication and Characterization of Zhejiang Province, School of Engineering, Westlake University, 18 Shilongshan Road, Hangzhou 310024, Zhejiang Province, China}
\affiliation{Institute of Advanced Technology, Westlake Institute for Advanced Study, 18 Shilongshan Road, Hangzhou 310024, Zhejiang Province, China}

\author{Wei Yan}
\affiliation{Key Laboratory of 3D Micro/Nano Fabrication and Characterization of Zhejiang Province, School of Engineering, Westlake University, 18 Shilongshan Road, Hangzhou 310024, Zhejiang Province, China}
\affiliation{Institute of Advanced Technology, Westlake Institute for Advanced Study, 18 Shilongshan Road, Hangzhou 310024, Zhejiang Province, China}

\author{Min Qiu}
\thanks{ qiu\_lab@westlake.edu.cn }
\affiliation{Key Laboratory of 3D Micro/Nano Fabrication and Characterization of Zhejiang Province, School of Engineering, Westlake University, 18 Shilongshan Road, Hangzhou 310024, Zhejiang Province, China}
\affiliation{Institute of Advanced Technology, Westlake Institute for Advanced Study, 18 Shilongshan Road, Hangzhou 310024, Zhejiang Province, China}

\begin{abstract}

The transfer of angular momentum carried by photons into a microobject has been widely exploited to achieve the actuation of the microobject.
However, this scheme is fundamentally defective in nonliquid environments as a result of the scale gap between friction forces ($\mu$N) and optical forces (pN).
To bypass this challenge, the researchers have recently proposed to take advantage of elastic waves based on opto-thermo-mechanical effects~\cite{Lu:2017,Lu:2019,Linghu:2021,Tang:2021}.
Grounded on this insight, we here demonstrate and characterize the in-plane rotation of a gold nanoplate in its surface contacting with a microfiber, driven by nanosecond laser pulses, which has not been explored before.
Furthermore, we examine the underlying physical mechanisms and highlight the essential role of the spatial gradient of optical absorption.
The combined experimental and theoretical results offer new insights into the study of the light-induced actuation of the microobjects in nonliquid environments, an emerging field far from being mature in both comprehensive understanding and practical applications.
\end{abstract}
\maketitle

\section*{Introduction}
Due to its unique advantages of precision, immediacy and miniaturization, manipulation of micro- and nano-objects with optical forces has proved to be an indispensable tool in a myriad of applications~\cite{Grier:2003}, such as biological manipulation and detection~\cite{Nadappuram:2019}, micro-flow control~\cite{MacDonald:2003}, and particle delivery~\cite{Tomas:2005}. This powerful technique is based on a straightforward principle: light waves carry momentum that can transfer to objects during scattering and absorbing processes, and, accordingly, enable their motions~\cite{Ashkin:1970,Ashkin:1986,Gao:2017,Tanaka:2020}. Specifically, the transfer of linear momentum results in pushing or pulling forces that are widely used for optical trapping~\cite{Chen:2011,Marago:2013}, while the transfer of angular momentum induces mechanical torques causing objects to rotate~\cite{He:1995,Curtis:2003,Neale:2005,Liu:2010}. Despite its great success, the state-of-the-art optical manipulation is largely limited in liquid environments, and inevitably fails in nonliquid environments due to the magnitude mismatch between optical forces (pN) and interfacial friction (adhesive) forces ($\mu$N).

To break down the barrier and realize actuation of microobjects in nonliquid environments, a few recent studies by us and others exploited elastic waves---which, like light waves, also carry momentum---based on opto-thermo-mechanical effects~\cite{Lu:2017,Lu:2019,Linghu:2021,Tang:2021}. These studies employed optical microfiber-based systems---that were initially proposed by Lu et al.~\cite{Lu:2017}---and demonstrated various motion patterns of microobjects on the dry surfaces of microfibers. Specifically, microobjects were observed to move around the microfibers along the azimuthal/axial directions, when nanosecond laser pulses are delivered into the microfibers. The motion directions are controllable by adjusting relative positions and contact configurations between the microobjects and the microfibers. However, among all these existing observations, an important degree of freedom of motion---rotation of microobjects in their surfaces that contact with the fixed supports, hereafter termed as {\it in-plane rotation}---remains to be comprehensively investigated.

In this article, we fill this gap by experimentally demonstrating the light-induced in-plane rotation of gold plates on microfibers, and theoretically discussing its underlying mechanisms.
The observed in-plane rotation is driven by laser pulses in a stepwise fashion.
Besides, the initiation and termination of the in-plane rotation is observed to be correlated with the symmetry configurations of the two wings of the gold plates.
Comparing with the existing optical-force-based approaches that enable the in-plane rotation by transferring either the orbital angular momentum (OAM) or the spin angular momentum (SAM) of light to objects, our approach shows three {\it complementary significances}. First, it works in nonliquid environments based on the opto-thermo-mechanical effects, while the other optical-force-based approaches typically apply in liquid environments. Second, through guiding light efficiently along microfibers, our approach is free of complex light paths, which is thus easy to be implemented. Third, the theoretical analysis suggests that the in-plane rotation could be potentially controlled by spatial profiles of optical absorption, thereby bringing a new means for optical manipulation. Our proposed light-induced in-plane rotation of microobjects thus contributes a new manipulable degree of freedom of motion of microobjects in nonliquid environments, which is expected to constitute a basis for future developments of microscopic mechanical actuators.

\section*{Observation of in-plane rotation}

\begin{figure*}[th!]
\includegraphics[width=0.87\textwidth]{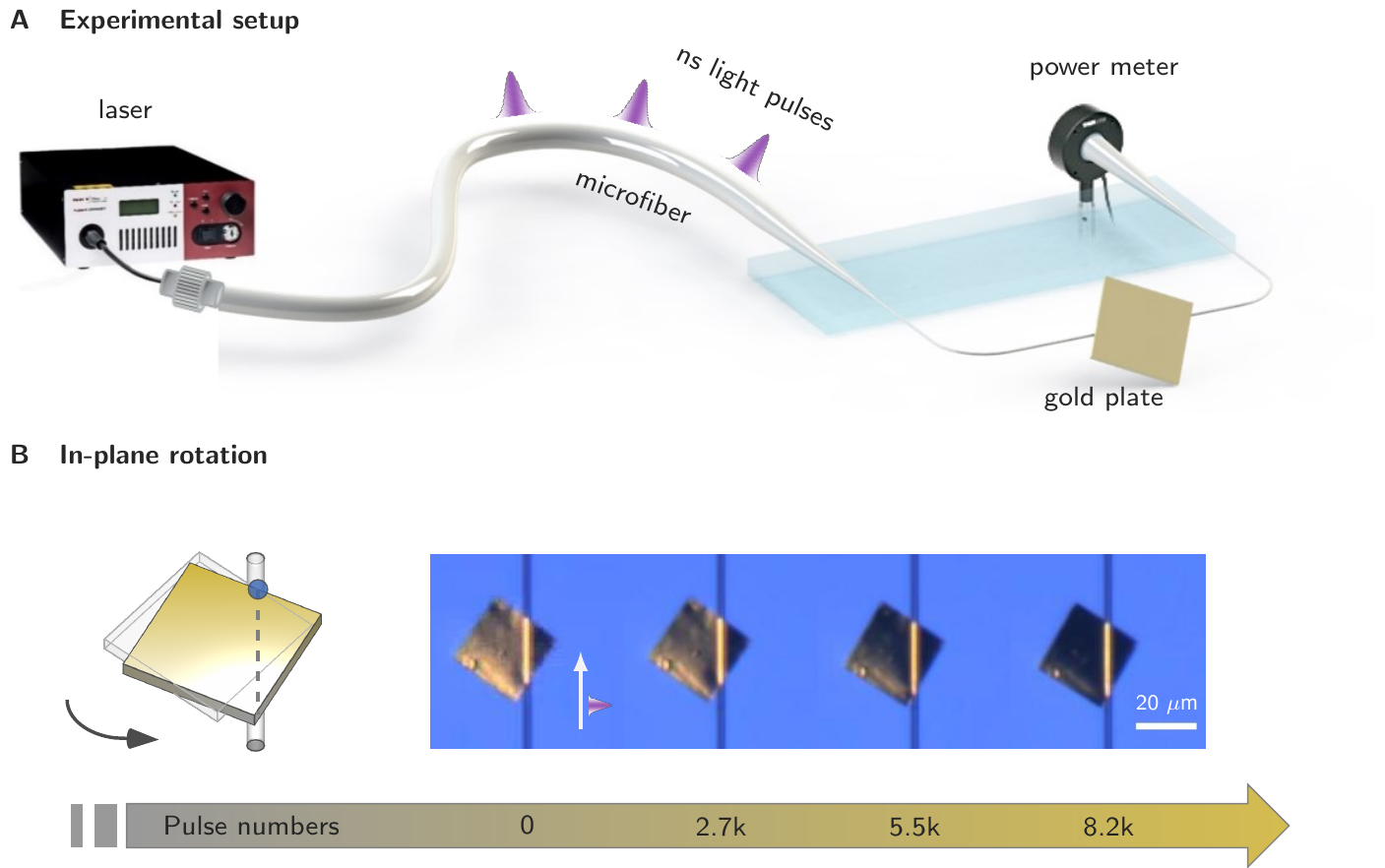}
\caption{{\bf Light-induced in-plane rotation of gold plates on silica microfibers}.
{\bf A}. Sketch of the experimental setup, involving a nanosecond-pulsed supercontinuum laser source (pulse duration, 2.6 ns; repetition rate, 230 Hz; wavelength range, 450-2400 nm; average power, 0.23 mW), a microfiber with 2-$\mu$m diameter, and a square gold plate (side length, 30 $\mu$m; thickness, 60 nm) adhered to the microfiber.
{\bf B}. Sequential optical images recording the in-plane rotation of the gold plate.
}
\label{fig::1}
\end{figure*}

Figure ~\ref{fig::1}A sketches the experimental setup, which consists of three parts: a microfiber, a gold plate, and a supercontinuum light. The microfiber with diameter about a few micrometers was tapered from a standard multi-mode fiber~\cite{Tong:2013}. The gold plate was prepared by electron beam lithography (EBL; see Supplementary Fig. S1 for fabrication details). Compared to chemical methods~\cite{Guo:2006} that are widely used in synthesizing thin gold plates with predefined shapes, e.g., with hexagonal and triangular base shapes, the EBL is advantageous in fabricating nanostructures with user-designed shapes, such as square plates used in our experiments. The EBL fabricated gold plate was initially on a $\rm SiO_2$ substrate, and then, transferred onto the surface of the microfiber with a tapered fiber. To drive the locomotion of the gold plate, we used a nanosecond (ns) super-continuum light source, which covers a wavelength ranging from 450 nm to 2400 nm with a pulse duration of 2.6 ns. Notably, the employment of a ns pulsed laser source is essential in our actuation scheme, since it can efficiently generate pulsed elastic waves that carry necessary momentum to overcome friction resistances~\cite{Tang:2021}.

\begin{figure}[h!]
\includegraphics[width=0.47\textwidth]{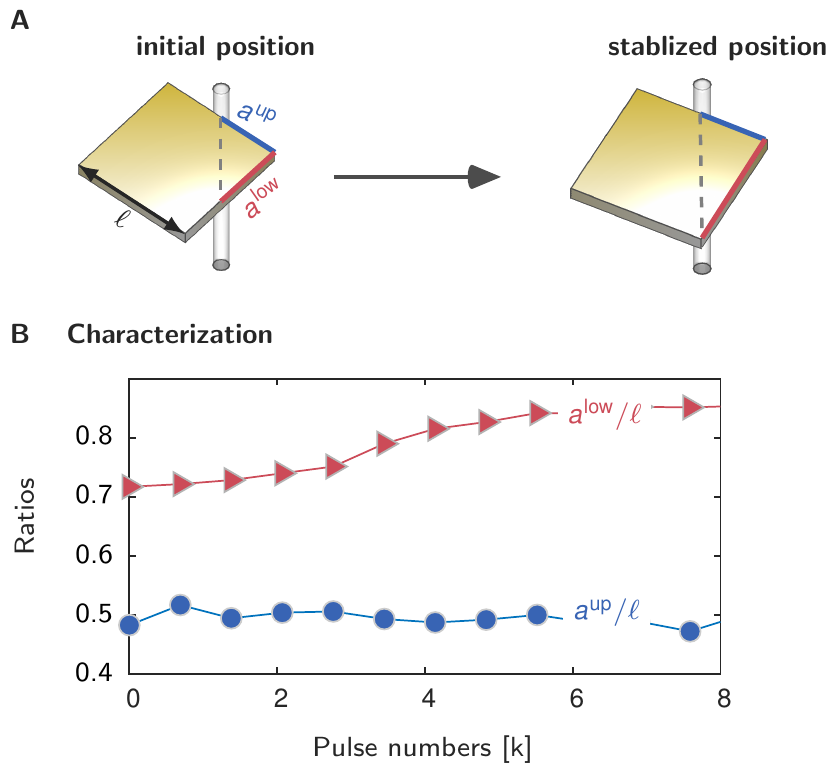}
\caption{{\bf Quantitative analysis of in-plane rotation}.
{\bf A}. Sketch of the in-plane rotation observed in Fig. 1. $a^{\rm up}$ and $a^{\rm low}$ represent the lengths of the right sections of the up and low edges of the plate, while $\ell$ represents the side length of the plate.
{\bf B}. Edge ratios, $a^{\rm up}/\ell$ and $a^{\rm low}/\ell$, as functions of the numbers of the delivered laser pulses.
}
\label{fig::2}
\end{figure}

\begin{figure}[ht!]
\includegraphics[width=0.47\textwidth]{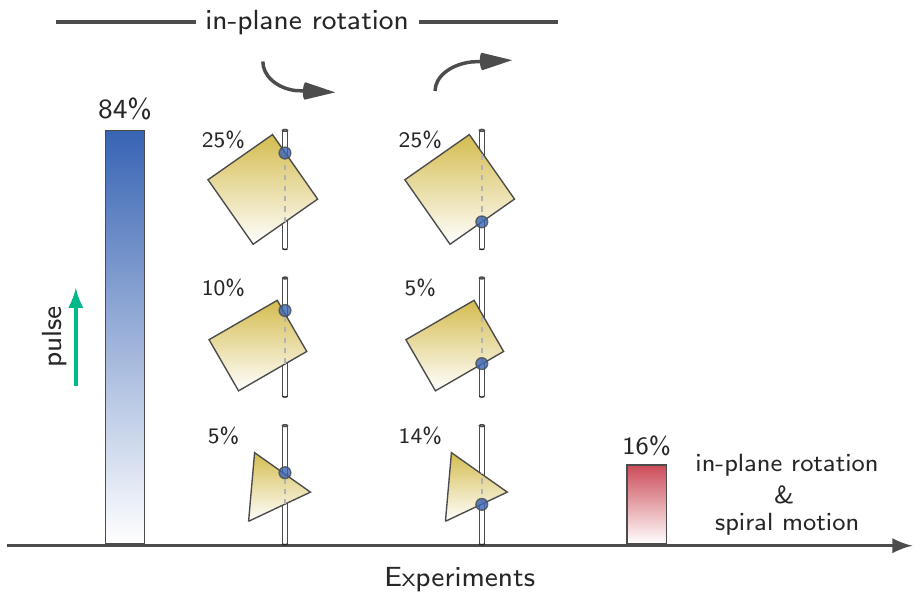}
\caption{{\bf Statistics summary of patterns of in-plane rotation involving 43 tests}. The blue dots represent the unrotated point, around which the in-plane rotation proceeds.
}
\label{fig::3}
\end{figure}

In this microfiber-based system, the previous work has demonstrated that the gold plate can move on the surface of the microfiber in both the azimuthal and axial directions~\cite{Lu:2017,Lu:2019,Linghu:2021,Tang:2021}. We here demonstrate a new motion pattern: the gold plate rotating in the plane of its lower surface that contacts with the microfiber, the so-called in-plane rotation. Figure~\ref{fig::1}B plots sequential optical images of a rotated gold square plate with side length of 30 $\mu$m and thickness of 60 nm. Initially, the gold plate is placed on the microfiber randomly. When a specific number of the laser pulses are delivered into the microfiber, the sequencing changes of the relative positions between the plate and the microfiber can be observed. The plate rotates counterclockwise with the upper endpoint of the touching line---between the plate and the microfiber---fixed. This process continues until the lower point of the touching line about approaches the bottom right vertex of the square plate. Finally, the plate stops its motion despite a continuous delivery of the laser pulses.
\begin{figure}[!b]
\begin{minipage}{1\textwidth}
\includegraphics[width=0.78\textwidth]{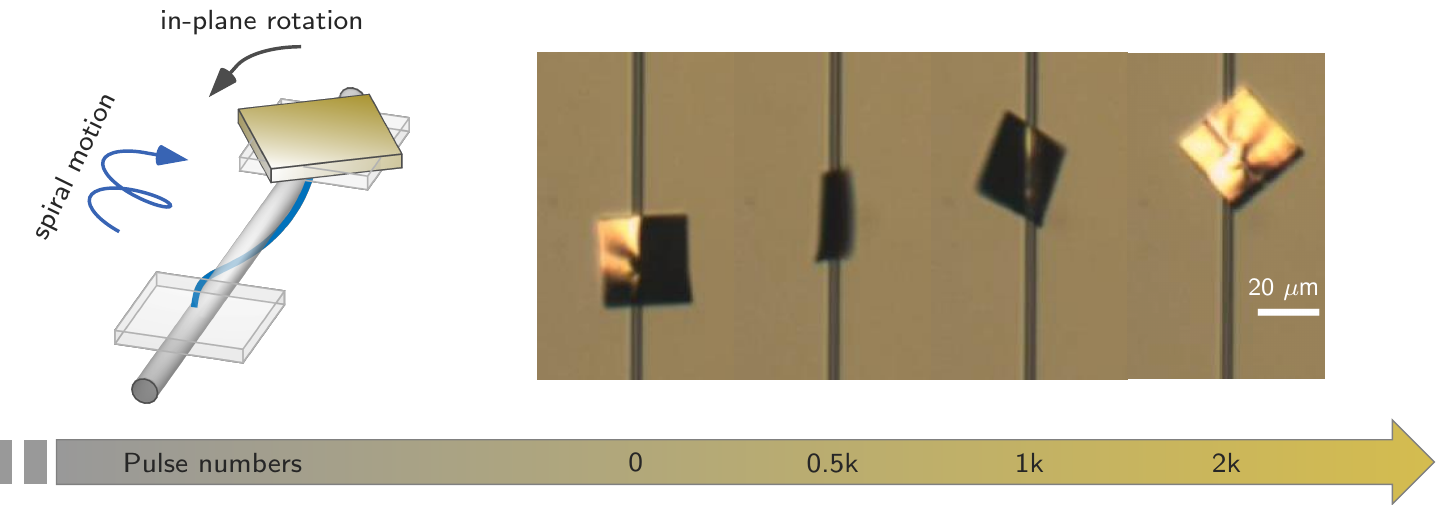}
\caption{{\bf Combined in-plane rotation and spiral motion of gold plates on silica microfibers}. (Left) Sketch of the observed motion. The gold plate moves in both the azimuthal and axial directions of the microfiber, resulting in the spiral motion. At the same time, the gold plate changes its relative position with respect to the microfiber due to the in-plane rotation. (Right) Sequential optical images recording the motion.
}
\label{fig::4}
\end{minipage}
\end{figure}

We characterize the in-plane rotation by monitoring the changes of the ratios associated with the right two edges of the plate, $a^{\rm up}/\ell$ and $a^{\rm low}/\ell$ (see their definitions in Fig. ~\ref{fig::2}A). As shown in Fig. ~\ref{fig::2}B, the upper-edge ratio $a^{\rm up}/\ell$ keeps a constant value about 0.5, quantifying that the upper point of the touching line is almost unrotated during the in-plane rotation, while the lower-edge ratio $a^{\rm low}/\ell$  gradually increases from 0.7 to a stabilized value about 0.85. This process takes a total number of the laser pulses to about 6k. Accordingly, the averaged rotation angle of the plate per single laser pulse is estimated to be about $10^{-3}$ degree, implying that the lower endpoint of the touching line slides about 1 nm per single laser pulse on average.
\noindent{}\\ \\

\section*{Regular Pattern of in-plane rotation}

To infer the regular pattern of the in-plane rotation, we performed additional experiments involving gold plates with different base shapes (square, rectangle, and triangle; see Supplementary Fig. S2). As summarized up in Fig.~\ref{fig::3}, out of 43 tests that show the in-plane rotation, most ($\sim$84\%) are similar as Fig.~\ref{fig::1}. The plates rotate about one endpoint of the touching line. The unrotated endpoint could locate on either edge of the plate that intersects with the touching line (labelled with blue circles in Fig. \ref{fig::3}), which is found to be independent of the direction of the laser incidence. The plates exhibit the same trend of their final positions after the in-plane rotation,  that is, one of their vertices approaches the microfiber, as shown in Fig. \ref{fig::1} and Supplementary Fig. S2.

The other minority tests ($\sim$16\%) show that the in-plane rotation could take place along with the spiral motion. The spiral motion implies that the plate revolves around the axis of the microfiber while simultaneously translating parallel to the axis, which has been comprehensively studied in Ref.~\cite{Tang:2021}. Figure~\ref{fig::4} plots the sequential optical images recording the combined in-plane rotation and spiral motion of a gold plate. With the in-plane rotation, the plate continuously changes its position relative to the microfiber, finally reaching a stable position with one of the vertices approaching the microfiber, after the delivery of 2k laser pulses. Then, the plate terminates the in-plane rotation, while the stable spiral motion remaining. Note that, differing from Fig. ~\ref{fig::1}, the central point of the in-plane rotation is now on the middle of the touching line, instead of the endpoint.

Despite definite differences between individual cases, all these tests exhibit the common feature: the in-plane rotation is always towards the direction that intends to reduce the area difference between the two wings of the gold plate, i.e., making the two wings become more symmetric. This observation implies that the symmetry breaking of the two wings should be a necessity for triggering the in-plane rotation, which shall be further examined with numerical studies.
\\
\newpage

\section*{Theoretical Analysis }
As revealed in Refs.~\cite{Lu:2019,Tang:2021}, in the present system, elastic waves induced by optical absorption in the gold plate can provide sufficiently mechanical momentum to overcome $\mu$N-scale friction forces and enable the locomotion of the plate. Based on this insight, it is reasonable to expect that the physical mechanism underlying the observed in-plane rotation also associates with the laser-induced elastic waves.

\begin{figure*}[ht!]
\includegraphics[width=0.95\textwidth]{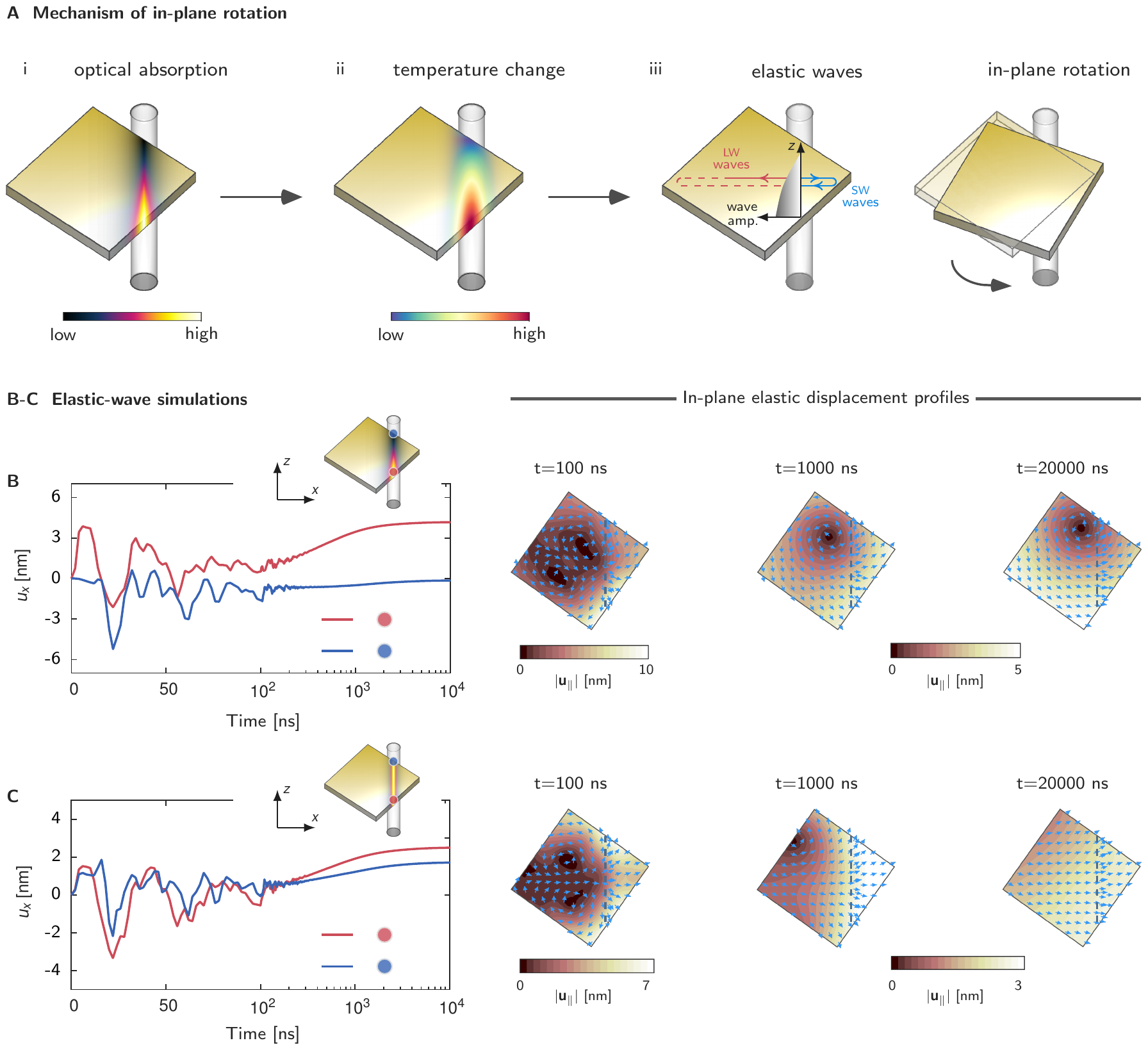}
\caption{{\bf Proposed mechanism of light-induced in-plane rotation of gold plates on silica microfibers}. {\bf A}. Sketch of three-step physical processes involved in the in-plane rotation. Step i: Laser pulses induce a gradient distribution of the optical absorption in the gold plate along the touching line between the plate and the microfiber. Step ii: The gradient distribution of the optical absorption translates its feature to the temperature distribution. Step iii: The temperature changes excite elastic waves, whose local amplitudes are (approximately) proportional to the local temperature values; consequently, the elastic displacements along the touching line follow the gradient distribution, resulting in the in-plane rotation.
{\bf B-C}. Elastic-wave simulations considering two extremes in the spatial distribution of the optical absorption: a linear increment from the upper endpoint to the lower endpoint of the touching line ({\bf B}) and a uniform distribution along the touching line ({\bf C}). (Left) Evolution of the elastic displacements along the azimuthal direction of the microfiber (i.e., the $x$ direction) for the two endpoints (marked with red and blue circles in the insets) of the touching line as driven by a single laser pulse. (Right) Profiles of in-plane elastic displacements at different times in the $y-z$ plane through the center of the gold plate at different time, t=100, 1000, 20000 ns. The parameters of the gold plate, the microfiber, and the laser pulse (pulse energy, 0.8 nJ) are adopted from the experimental values in Fig.~\ref{fig::1}, and the position between the gold plate and the microfiber is set to be the initial position pre-in-plane-rotation. The absorbed optical power has a Gaussian profile with a width of 500 nm in the $x$-direction, and distributes uniformly across the thickness of the plate. The life time of elastic waves is set to be 40 ns.
}
\label{fig::6}
\end{figure*}

Briefly, the involved physics features opto-thermal-elastic processes, which can be subdivided into three steps (see Fig.~\ref{fig::6}A). First, as laser pulses propagate in the microfiber, their electric fields leak out and are absorbed by the plate. Second, the absorbed optical power is converted into heat, which results in the temperature rising of the plate, and, then, induces the coherent lattice oscillations, i.e., elastic waves. Third, the induced elastic waves, bouncing back and forth in the plate, lead to the motion of the plate. Below, we analyze them in detail, and focus on numerically reproducing the observed in-plane rotation for the exemplified case of Fig.~\ref{fig::1}.
\\

\begin{figure}[h!]
\includegraphics[width=0.45\textwidth]{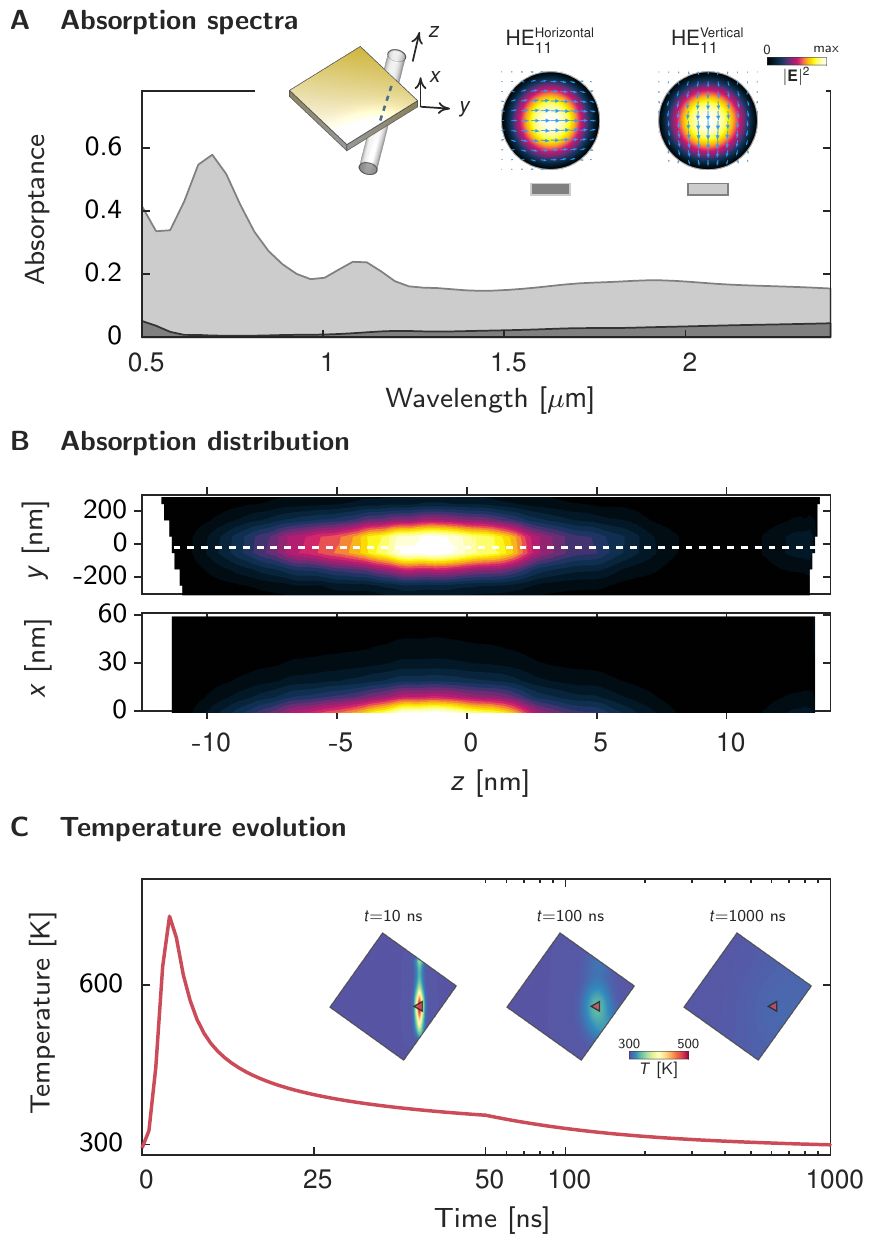}
\caption{{\bf Opto-thermal simulations of a gold plate and microfiber coupled system}. {\bf A}. Absorption spectra of the system under the incidences of two fundamental modes of the microfiber, $\rm HE_{11}^{\rm \scriptscriptstyle Vertical}$ and $\rm HE_{11}^{\rm \scriptscriptstyle Horizontal}$ (see the insets for their profiles). {\bf B}. Distributions of electric field intensities $|\mathbf E|^2$ on the lower surface ($y-z$ plane; top panel) and in the sliced $x-z$ plane (through the touching line; bottom panel) of the gold plate, at the absorption peak wavelength (700 nm) under the incidence of the $\rm HE_{11}^{\rm \scriptscriptstyle Vertical}$ mode. {\bf C}. Characteristic temperature evolution of the system at a selected point (marked by triangles in the insets) in the gold plate as driven by a single laser pulse. Insets: temperature distributions in the $y-z$ plane through the center of the gold plate at different time, $t$=10, 100, 1000 ns. The parameters of the gold plate, the microfiber, and the laser pulse (pulse energy, 0.8 nJ) are adopted from the experimental values in Fig. ~\ref{fig::1}, and the position between the plate and the microfiber is set to be the initial position pre-in-plane-rotation. The distribution of the absorbed optical power in {\bf B} is employed to compute the temperature evolution in {\bf C}.
}
\label{fig::5}
\end{figure}

\noindent{\bf Opto-Thermal Process }\\
The generation efficiency of the elastic waves relies on the opto-thermal conversion in the microfiber and gold plate coupled system. Here, a square gold plate placed on a microfiber---with the same geometrical parameters as in Fig.~\ref{fig::1}---is considered to simulate the involved opto-thermal process. The relative position between the plate and the microfiber is set to be the initial position pre-in-plane-rotation. Figure~\ref{fig::5}A plots the simulated absorption spectra of the system with two fundamental modes of the microfiber, $\rm HE_{11}^{\rm \scriptscriptstyle Vertical}$ and $\rm HE_{11}^{\rm \scriptscriptstyle Horizontal}$ (see the insets in Fig.~\ref{fig::5}A for their modal profiles), as incident waves. It is observed that the absorption efficiency for the $\rm HE_{11}^{\rm \scriptscriptstyle Vertical}$ incidence is much higher than that for the $\rm HE_{11}^{\rm \scriptscriptstyle Horizontal}$ incidence. Remarkably, the former shows a peaked absorptance close to 60\% at the incident wavelength of 700 nm. This is because that the $\rm HE_{11}^{\rm \scriptscriptstyle Vertical}$ mode has a large electric field component perpendicular to the surface of the gold plate, thereby favoring the plasmon excitations and enhancing the absorption.

Further, to simulate the temperature evolution of the system, the absorbed optical power is temporally modelled as a Gaussian pulse with 2.6-ns pulse width and 0.8-nJ pulse energy, and spatially distributed as at the absorption peak wavelength (700 nm) under the $\rm HE_{11}^{\rm \scriptscriptstyle Vertical}$ incidence. As shown in Fig.~\ref{fig::5}B, the absorbed optical power strongly localizes in a line-shaped sub-wavelength region around the touching line (marked with dashed lines). Moreover, the absorbed optical power is apparently inhomogeneous and shows a hot spot. Figure~\ref{fig::5}C plots the simulated temperature evolution at a representative point in the plate, which shows a rapid heating within the period of the incident laser pulse and then follows a rather slow cooling. Besides, the temperature distribution, at an earlier time before the thermal diffusion and cooling are proceeded sufficiently (e.g., t=10 ns, see the inset in Figure~\ref{fig::5}C), demonstrates a similar spatial profile as the optical absorption.
\\

\noindent{\bf Thermal-Elastic Process and Elastic-Wave Dynamics  }\\
The induced temperature changes centralized around the touching line excite the elastic waves in the plate. For convenience of discussion, the elastic waves are classified into “long-wing” (LW) and “short-wing” (SW) waves, according to their initial propagation directions towards either the LW or SW sides of the plate (see Fig.~\ref{fig::6}A-iii). As discussed in Ref.~\cite{Tang:2021}, albeit the reversions of their propagation directions due to reflections, these LW and SW waves retain the same oscillation directions that point towards the LW and SW sides, respectively. Consequently, as the LW and SW waves pass through the touching line between the plate and the microfiber, they consistently drive the plate to slide in the opposite directions (that is, towards the LW and SW sides, respectively). This process continues until the elastic waves are finally attenuated. The net motion direction of the plate points towards the SW side, since the SW waves are less attenuated thanks to their shorter traveling distances. This physical picture is consistent with our observation that, during the in-plane rotation, the averaged motion direction of the plate indeed points towards the SW side (see Fig.~\ref{fig::1} and Fig.~\ref{fig::4}), manifesting in the decrement of the area difference between two wings.

Apparently, if the elastic displacements along the azimuthal direction of the microfiber are uniform along the touching line, the plate shall revolve stably around the axis of the microfiber~\cite{Lu:2019} instead of the observed in-plane rotation. This recognition directly implies that, for the in-plane rotation, the azimuthal displacements should be extremely non-uniform with a distinct gradient feature. Referring to Fig.~\ref{fig::1}, such azimuthal displacements are observed to linearly grow from the upper endpoint of the touching line—that is fixed during the in-plane rotation—to the lower endpoint. Consequently, the amplitudes of the elastic waves should similarly follow a gradient distribution along the touching line. Besides, as mentioned above, the elastic waves are excited due to the temperature change that results from the optical absorption. Gathering these facts, it is nature to {\it conjecture that the non-uniform (gradient) distribution of the absorbed optical power along the touching line could be a possible physical mechanism that explains the observed in-plane rotation} (see the sketch in Fig.~\ref{fig::6}A).

To justify the proposed conjecture, we perform the thermal-elastic coupled simulations with COMSOL Multiphysics. The simulated system in Fig.~\ref{fig::6} models the experiment of Fig.~\ref{fig::1}. Moreover, to reveal the role of the spatial gradient of the optical absorption, we consider two extremes in the spatial distribution of the absorbed optical power. First, in Fig.~\ref{fig::6}B, the absorbed optical power is set to vary linearly along the touching line with the intensity zero on the upper endpoint and the intensity maximum on the lower endpoint. The left panel of Fig.~\ref{fig::6}B plots the temporal evolutions of the azimuthal displacements of the two endpoints on the touching line as driven by a single laser pulse. The results show that the stabilized displacement of the upper point (close to zero) is strikingly smaller than that of the lower point ($\sim$4 nm). This observation decisively confirms our intuitive expectation that the original gradient of the optical absorption can translate into the elastic displacements through the three-step opto-thermal-elastic processes (see Fig.~\ref{fig::6}A). The right panel of Fig.~\ref{fig::6}B visualizes the spatial profiles of the elastic displacements of the plate at different time ($t$=100, 1000, 20000 ns), demonstrating the in-plane rotation of the plate closely around the upper endpoint of the touching line.

Next, in Fig.~\ref{fig::6}C, we turn off the gradient feature by setting the absorbed optical power to be uniform along the touching line. The simulated results show that the azimuthal displacements of the two endpoints on the touching line become close to each other, and, thus, the in-plane rotation is significantly suppressed. The striking contrast between Figs.~\ref{fig::6}B and C thus validates our conjecture that the non-uniform (gradient) distribution of the optical absorption could potentially enable the observed in-plane rotation.

As the in-plane rotation proceeds, the two wings of the plate become more symmetric, as well as the more balanced elastic waves. Therefore, the azimuthal displacements of the plate should be gradually reduced, and, finally, the in-plane rotation ceases. We simulate this process by fixing the upper endpoint of the touching line, while varying the lower endpoint. The linear distribution of the absorbed optical power similar as in Fig.~\ref{fig::6}B is adopted. For different positions of the lower endpoint, the integrated total absorbed optical power is kept the same. The results plotted in Supplementary Fig. S3 show that, for the stabilized azimuthal displacements of the two endpoints on the touching line, their difference decreases and approaches zero, as the lower endpoint approaches the bottom right vertex of the square plate and the two wings become more symmetric.
\\

\noindent{\bf Remarks on theoretical interpretation }\\
The numerical results in Fig.~\ref{fig::6}, featuring the fundamental role of the gradient distribution of the optical absorption, brings out an interpretation for the observed in-plane rotation. As is shown in Fig.~\ref{fig::5}B, the electromagnetic interactions between the microfiber and the plate could possibly provide this required gradient. Moreover, as is well known, the plate is easily curved when transferred onto the microfiber~\cite{Tang:2021}, so that the optical absorption might pile up in the contacting part, while leaving the non-contacting part with minimum absorption. Therefore, the unevenness of the plate can be another important reason for generating the gradient distribution of the optical absorption. This possibly explains why, in Fig.~\ref{fig::3}, the specific location of the unrotated point (i.e., the absorption minimum) on the touching line shows no obvious dependence on the direction of the laser incidence.

Additionally, in the experiments, we observe that the plates generally cease the in-plane rotation as one of their vertices approaching the microfiber. This phenomenon cannot be explained with our numerical simulations, wherein the surface curvature of the plates is not taken into account. We guess that it might be due to that the plates could be more easily curved around their vertices, which leads to a drastic decreasing in the absorption power as the plates rotate to the vertex position, and, correspondingly, to the termination of the in-plane rotation.


\section*{Conclusions}
In this article, we demonstrate the in-plane rotation of gold plates on microfibers as driven by the nanosecond laser pulses. The actuation is realized by thermally excited elastic waves generated in the gold plates through the pulsed optical absorption. Moreover, we propose an interpretation for the in-plane rotation, featuring the key role played by the gradient distribution of optical absorption. Combining the experimental observations and the theoretical interpretation, we envision that detecting the in-plane rotation might provide a new way to infer the distribution of the optical absorption. On the other hand, by exploring various spatial profiles of the absorbed optical power, one can in principle be able to control the trajectory of the in-plane motion.
In conclusion, our results exemplify the promising capacity of elastic waves in driving locomotion of microobjects in nonliquid environments, which otherwise is formidable with optical-force-based approaches.
\\
\\
\\

\noindent{\bf Acknowledgements---}The authors acknowledge the support from the National Key Research and Development Program of China (Grant No. 2017YFA0205700), the National Natural Science Foundation of China (Grant No. 61927820, No. 61905201), and the China Postdoctoral Science Foundation (2020M671809).
\bibliography{References}
\end{document}